\documentclass[%
 reprint,
 amsmath,amssymb,
 aps,
]{revtex4-2}

\usepackage{graphicx}
\usepackage{dcolumn}
\usepackage{bm}

\usepackage{color}

\begin{document}

\title{Interaction of intense ultrashort laser pulses with solid targets: A systematic analysis using first-principles calculations}

\author{Atsushi Yamada}
\affiliation{Department of Applied Chemistry, National Defense Academy, Yokosuka 239-8686, Japan}
\author{Kazuhiro Yabana}
\affiliation{Center for Computational Sciences, University of Tsukuba, Tsukuba 305-8577, Japan}

\date{\today}

\begin{abstract}
Intense ultrashort laser pulse irradiation of solid targets was systematically investigated at the first-principles level, both theoretically and computationally.
In the method, the propagation of a pulsed light through a thin film is described by a one-dimensional Maxwell's equation, and the microscopic electronic motion at different positions in the film is described by employing first-principles time-dependent density functional theory (TDDFT).
The method uses a coarse-graining approximation to couple light propagation and electronic motion, and is termed the multiscale Maxwell-TDDFT method.
The reflectance, transmittance, and absorbance of pulsed light incident normally on thin films of 50--200 nm thickness were calculated for materials with different optical properties, such as aluminum (simple metal), graphite (semi-metal), silicon (small-gap dielectric), and quartz (wide-gap dielectric).
Optical response transitions were explored as the light intensity shifted from the linear regime, represented by the dielectric function for weak light, to the extremely nonlinear regime, represented by plasma reflection under intense light conditions. 
Numerous mechanisms that depend on the laser pulse intensity and material type were found to contribute to these changes. These include multiphoton absorption, saturable absorption, sign change of the effective dielectric constant, and transition from quantum occupation to classical Boltzmann distribution.
Thus, the calculations provide a unified understanding of the interaction of intense pulsed light with solids, occurring on an extremely short time scale.
\end{abstract}
\maketitle

\section{Introduction}

The interaction of light with solids varies significantly depending on various factors, such as frequency, intensity, polarization, and pulse duration of the light and the electronic structure of solids, dielectrics, or metals. These factors determine the characteristics of the light-matter interaction.
To theoretically describe the interaction of intense ultrashort pulsed light with various materials, it is important to clarify the basic physical pictures that change drastically depending on the temporal duration and intensity of the pulsed light.

We first considered the variation in the optical response as the light intensity increased.
For ordinary weak light, the dielectric function given by the perturbation theory of quantum mechanics determines the light-matter interactions. 
For moderately intense pulses, perturbative nonlinear optics phenomena such as the optical Kerr effect and second- and third-harmonic generation are extensively treated in nonlinear optics textbooks \cite{Boyd2008, Shen2003}.
As the intensity of pulsed light approaches the damage threshold of materials (but is still below the threshold), nonlinear responses, which cannot be described by the perturbation theory, become significant, and they have been extensively explored in the last decade.
For example, high harmonic generation from solids \cite{Ghimire2011, Luu2015, Ghimire2019} has attracted significant interest, possibly leading to the development of compact XUV sources.
Ultrafast changes in the optical properties of materials, such as the metallization of dielectrics \cite{Schiffrin2013}, have also attracted interest in the development of new optical device principles, such as petahertz optoelectronics \cite{Krausz2014}.
Above a certain critical intensity, materials suffer permanent and irreversible damage.
The formation of periodic microstructures at surfaces has been observed and attracting interest in the mechanism of ultrafast energy deposition \cite{Bonse2017}.
Femtosecond-laser processing using ultrashort laser pulses has attracted considerable interest as a tool for precise material processing \cite{Balling2013}.
Further intense laser pulses instantaneously created plasma on the surface of materials.
Strongly excited matter on a surface is characterized as warm dense matter \cite{Koenig2005}, which is an intermediate state between ordinary condensed matter and hot plasma.

Next, we considered the variation in the optical response as the temporal duration of the pulse decreased. 
When the pulse duration is sufficiently longer than a certain relaxation time, the irradiated material can be considered to be in a quasi-equilibrium state with respect to the related relaxation dynamics. 
Thus, there is a balance between excitation by the laser pulse and de-excitation by the relaxation process.
As the pulse duration decreases and approaches the relaxation time, the system is considered to be in a nonequilibrium state.
On further decreasing the pulse duration, if the pulse duration is sufficiently shorter than any kind of relaxation, a ballistic picture becomes appropriate for electronic motion. This entails treating them as independent electrons moving at an average one-body potential.

There have been numerous reports on systematic measurements of the reflectance, transmittance, and absorbance of intense ultrashort pulsed light from the surface or thin films of various solids \cite{Hulin1984, Milchberg1988, Fedosejevs1990, Price1995, Riley1998, Sokolowski2000, Eidmann2000, Chen2001, Fisher2001, Sabbah2002, Ziener2002, Rajeev2005, Ping2006, Cerchez2008, Breusing2009, Winzer2012, Danilov2015, Meng2016, Blumenstein2020, Genieys2020, Genieys2021}.
For example, in [\onlinecite{Price1995}], measurements were reported using laser pulses with a wavelength of 800 nm, a duration of 120 fs, and intensities ranging $10^{13} \sim 10^{18}$ W/cm$^2$ for materials such as wide-gap dielectrics, quartz, and metals (Al, Au, etc.).
While all materials eventually reflected the incident pulse by plasma reflection at intensities above $10^{15}$ W/cm$^2$, a strong absorption was observed below this intensity, with the strongest absorption observed in quartz.
Numerous theoretical approaches have been developed to describe the carrier dynamics and light propagation in solids. In metals, the increase in electronic temperature due to laser irradiation and resulting change in the dielectric constant have been considered \cite{Eidmann2000, Fisher2001, Suslova2017, Kim2023}. 
In dielectrics, generation of free carriers is the most significant process, and it is described by multiple rate equations \cite{Sokolowski2000, Kaiser2000, Rethfeld2006, Petrov2008, Medvedev2010, Rethfeld2017}. 
In either case, the carrier density and collision frequency of the electrons are the decisive parameters that determine the propagation of the intense laser pulses within solids.

In this paper, we report a systematic theoretical investigation of the reflectance, transmittance, and absorbance of the intense ultrashort pulsed light from thin films of various materials, using the first-principles computational method in materials science.
To obtain a unified view of the light-matter interaction in extremely short time scales for wide intensities, we performed systematic calculations of the propagation of a few-cycle pulsed light of optical frequency in thin films of various materials, including metals (aluminum), semi-metals (graphite), small-gap dielectrics (silicon), and wide-gap dielectrics ($\alpha$-quartz).
The calculated results reveal the basic features of the optical response, including a linear response for a weakly pulsed light, where the dielectric function determines the response, and an extremely nonlinear response for an intense pulsed light, where plasma reflection eventually dominates.

We employed multiscale formalism to describe linear and nonlinear light propagation.
At the macroscopic scale, we solved a one-dimensional Maxwell's equation to calculate the reflectance, transmittance, and absorbance of the pulsed light that normally irradiate thin films of a wide variety of materials.
At the microscopic scale, we employed the first-principles time-dependent density functional theory (TDDFT) \cite{Runge1984, Ullrich2012} to calculate the electronic motion in a unit cell of crystalline solids at each grid point in macroscopic coordinates.
By solving the time-dependent Kohn--Sham (TDKS) equation in real time with the electric field of the pulsed light as an external potential, the Bloch orbitals can be evolved in time to obtain the electric current induced by the pulsed light \cite{Bertsch2000, Otobe2008}.
By simultaneously solving the Maxwell and TDKS equations, we can self-consistently describe the light propagation and electronic motion.
We call this framework the multiscale Maxwell-TDDFT method \cite{Yabana2012}.
For weakly pulsed light, this calculation is equivalent to the description of light propagation by macroscopic electromagnetism, where a dielectric function given by the TDDFT is utilized.
For an intense pulsed light, this method can describe nonlinear light propagation without any perturbation theory expansions. 

The remainder of this paper is organized as follows. 
In Section II, we present theoretical and computational methods to describe nonlinear light propagation as well as first-principles light-electron interactions. 
In Section III, the results of systematic calculations are presented and discussed. 
Section IV presents a summary of the results.

\section{Theoretical and computational method \label{sec:method}}

We investigated nonlinear light propagation through thin films of four typical materials, namely aluminum (Al), graphite, silicon (Si), and $\alpha$-quartz (SiO$_2$), with different optical properties.
For a thin film of thickness 50--200 nm placed in a vacuum, a linearly polarized, few-cycle pulsed light irradiates normally on the film.
The multiscale Maxwell-TDDFT method developed in Ref.~[\onlinecite{Yabana2012}] is used to describe light propagation and electronic motion inside the thin film.
In this section, we first describe the formalism based on the TDDFT for calculating the electronic motion in a unit cell of a crystalline solid, driven by an external, spatially-uniform electric field \cite{Bertsch2000, Otobe2008}.
We then explain the coupling of the TDDFT framework in a unit cell with a one-dimensional Maxwell's equation that describes light propagation.

\subsection{Electronic motion in a unit cell}

We describe the electronic motion in a unit cell of a crystalline solid driven by a spatially uniform pulsed electric field with a given time profile, using the orbitals $\left\{ u_{n{\bm k}}({\bm r},t) \right\}$ as functions of the spatial vector ${\bm r}$ and time $t$. Here, $n$ is the band index, and ${\bm k}$ is the crystalline momentum. The orbitals satisfy the TDKS equation, \cite{Bertsch2000,Otobe2008}
\begin{equation}
    i\hbar \frac{\partial}{\partial t} u_{n{\bm k}}({\bm r},t)
    =
    h_{\rm KS}\left[{\bm k} + {\frac{e}{\hbar c}\bm A}(t) \right] u_{n{\bm k}}({\bm r},t),
    \label{eq:tdks}
\end{equation}
with the Kohn-Sham Hamiltonian
\begin{eqnarray}
h_{\rm KS}[{\bm k}]    
    &=&  \frac{1}{2m} \left( -i\hbar \nabla + \hbar{\bm k} \right)^2 - e\phi({\bm r},t) 
\nonumber\\    
    &+&  \hat v_{\rm{NL}}^{\bm k} +
V_{\rm{xc}}({\bm r},t)\ . \quad
      \label{eq:hks}
\end{eqnarray}
Here, the vector potential ${\bm A}(t)$ is related to the applied electric field ${\bm E}(t)$ as follows:
\begin{equation}
{\bm E}(t) = - \frac{1}{c} \frac{d}{dt} {\bm A}(t).
\end{equation}
We treat the interaction of valence electrons with ions using a norm-conserving pseudopotential \cite{Troullier1991}.  
The scalar potential $\phi({\bm r},t)$ exhibits the same periodicity as the lattice 
and satisfies the following equation:
\begin{equation}
    \nabla^2 \phi({\bm r},t) = -4\pi e (n_{\rm ion}({\bm r}) - n_{\rm
      e}({\bm r},t) ) \ ,
\end{equation}
where the ionic charge density $n_{\rm ion}({\bm r})$ is included to produce the local part of the pseudopotential in $\phi$. 
In this study, the atomic positions are fixed during the time evolution calculations.
The electron density $n_{\rm e}({\bm r},t)$ is given by
\begin{equation}
    n_{\rm e}({\bm r},t) = \frac{1}{N_k}\sum_{n,{\bm k}}^{\rm occ}
    \vert u_{n{\bm k}}({\bm r},t) \vert^2,
\end{equation}
where $N_k$ denotes the number of ${\bm k}$-points.
The sum $n$ is taken over the occupied bands.
The nonlocal part of the pseudopotential is modified for a ${\bm k}$-point system as $\hat{v}_{{\rm NL}}^{{\bm k}}\equiv e^{-i{\bm k}\cdot{\bf r}}\hat{v}_{{\rm NL}}e^{i{\bm k}\cdot{\bm r}}$. Here, $\hat v_{\rm NL}$ is the nonlocal part of the pseudopotential for which we take a usual separable form \cite{Kleinman1982}. 
$V_{\rm xc}({\bf r},t)$ is the exchange-correlation potential, for which the adiabatic approximation is assumed based on either the local density approximation \cite{Perdew1981} or meta-GGA approximation \cite{Tran2009}.
The Bloch orbitals were initially set as ground-state solutions.

By solving the TDKS equation for a given electric field specified by the vector potential ${\bm A}(t)$, we obtain various physical quantities, including the electric current density, electronic excited energy, number density of excited electrons, and dielectric function after the pulse ends.
The quantities are defined as follows:

The electric current density averaged over the unit cell volume $\Omega$ is given by
\begin{eqnarray}
&&{\bm J}[{\bm A}(t)](t)=-\frac{e}{m}\int_{\Omega}\frac{d {\bm r}}{N_k \Omega}\sum_{n,{\bm k}}^{{\rm occ}}u_{n{\bm k}}^*({\bm r},t) \nonumber \\
&&\times \left\{-i\hbar\nabla+\hbar{\bm k}+\frac{e}{c}{\bm A}(t)+\frac{m}{i\hbar}\left[{\bm r},\hat{v}_{{\rm NL}}^{{{\bm k}+\frac{e}{\hbar c}{\bm A}(t)}}\right]\right\}u_{n{\bm k}}({\bm r},t).\nonumber \\
\label{eq:current}
\end{eqnarray}
This can be regarded as the macroscopic electric current for a given electric field of pulsed light and be utilized as the constitutive relationship in the multiscale Maxwell-TDDFT method.
For a sufficiently weak pulse, the calculated macroscopic electric current coincides with that given by the linear constitutive relationship with the conductivity tensor $\sigma_{\mu\nu}^0(t)$.
\begin{equation}
J_{\mu}(t) = \sum_{\nu} \int^t dt' \sigma^0_{\mu\nu}(t-t') E_{\nu}(t'),
\end{equation}
where $\mu,\nu$ denote the Cartesian components, $x,y,z$.
To indicate that this conductivity is related to the ground state, we used the superscript 0 in the conductivity.
This relationship is regarded as the definition of the TDDFT conductivity in the ground state of the material.
By solving the TDKS equation in real-time with a strong electric field, we can obtain a macroscopic electric current that includes the effects of nonlinear electronic motion without perturbative approximations.

We adopted the following definition for the number density of the excited electrons in the unit cell.
\begin{equation}
n_{\rm ex}(t) =\frac{1}{\Omega} \left\{ N_{\rm e} - \sum_{n n' \bm k} 
\left\vert \langle u^0_{n{\bm k+(e/c){\bm A}(t)}} \vert u_{n'{\bm k}}(t) \rangle \right\vert^2 \right\},
\label{eq:nex}
\end{equation}
where $N_e$ is the total number of electrons in the unit cell and $u^0_{n{\bm k}+(e/c){\bm A}(t)}$ is the Houston function, which is the Bloch wave function in the ground state with a shifted vector potential at time $t$.
It is also feasible to calculate the electronic excitation energy at time $t$ in the unit cell, which is given by the difference between the energy at time $t$ and that in the ground state.
A systematic calculation is reported in [\onlinecite{Yamada2019}].

Finally, we introduce a dielectric function at the end of the applied pulse.
To introduce this concept, we regard the applied pulsed electric field as a pump pulse and denote it as ${\bm E}^{\rm pump}(t)$. 
The current density induced by the pump pulse is expressed as ${\bm J}^{\rm pump}(t)$.
In addition to the pump pulse, a weak impulsive electric field is applied as the probe pulse.
\begin{equation}
{\bm E}^{\rm probe}(t) = -\frac{\Delta {\bm A}}{c} \delta(t-t_0),
\label{eq:probe}
\end{equation}
where $t_0$ denotes the time at which the pump pulse ended.
An impulsive electric field can be computationally applied using a small shift in the vector potential, ${\bm A}(t) \rightarrow {\bm A}(t) + \Delta {\bm A}\theta(t-t_0)$.
We denote the current density due to both pump and probe pulses as ${\bm J}^{\rm pump+probe}(t)$.
The conductivity of the electronic state excited by a pump pulse is introduced in \cite{Sato2014}
\begin{equation}
J^{\rm pump+probe}_\mu(t) - J^{\rm pump}_\mu(t) = \sum_\nu \int^{t} dt' \sigma^{\rm pump}_{\mu \nu}(t-t') E^{\rm probe}_\nu(t').
\end{equation}
Using the conductivity, we obtained the dielectric function after the pump pulse ended.
\begin{equation}
\epsilon^{\rm pump}_{\mu\nu}(\omega) = \delta_{\mu\nu} + \frac{4\pi i \sigma^{\rm pump}_{\mu\nu}(\omega)}{\omega},
\label{eq:eps_pump}
\end{equation}
where $\sigma^{\rm pump}_{\mu\nu}(\omega)$ is the Fourier transform of $\sigma^{\rm pump}_{\mu\nu}(t)$.
The defined dielectric function depends on $t_0$.
However, this dependence is weak \cite{Sato2014}.

\subsection{Multiscale Maxwell-TDDFT method}

Next, we explain the multiscale Maxwell-TDDFT method \cite{Yabana2012}, which describes light propagation by coupling the formalism of the microscopic electronic motion described above with the macroscopic one-dimensional Maxwell's equation.
We consider the irradiation of a thin film of thickness $d$ in vacuum by ultrashort pulsed light of a linearly polarized plane wave at normal incidence.
We assume that the thin film is in the $yz$-plane with polarization in the $z$-direction and that the pulsed light propagates along the $x$-axis.
For one-dimensional light propagation, Maxwell's equation yields a one-dimensional wave equation for the vector potential.
\begin{equation}
 \left (\frac{1}{c^2} \frac{\partial^2}{\partial t^2}  - \frac{\partial^2}{\partial X^2} \right) {\bm A}_{X}(t)
= \frac{4\pi}{c}   {\bm J}_{X}(t) \ ,
\label{eq:multiscale}
\end{equation}
where $X$ is the coordinate describing the light propagation. 
The material is assumed to be placed in the region $0 < X < d$.
${\bm A}_{X}(t)$ and ${\bm J}_{X}(t)$ are the vector potential and current density, respectively, at position $X$.  
To solve this equation, we introduce a uniform spatial grid for coordinate $X$.

To solve Eq.~(\ref{eq:multiscale}), we must determine ${\bm J}_{X}(t)$ at each grid point specified by $X$ inside the thin film for a given vector potential ${\bm A}_{X}(t)$.
For this purpose, we used the microscopic electronic motion described in the previous subsection.
At each point $X$ inside the film, we consider an infinitely extended electronic system under a time-dependent electric field ${\bm E}_X(t) = -(1/c)(d/dt){\bm A}_X(t)$. Here, the electric field ${\bm E}_X(t)$ is regarded as spatially uniform, and the dependence on microscopic coordinate ${\bm r}$ is ignored.
Then, the electronic motion at point $X$ is described by the Bloch orbitals $u_{n{\bm k},X}({\bm r},t)$ that satisfy the TDKS equation.
\begin{equation}
    i\hbar \frac{\partial}{\partial t} u_{n{\bm k},X}({\bm r},t)
    =
    h_{\rm KS} \left[{\bm k}+\frac{e}{\hbar c}{\bm A}_X(t) \right] u_{n{\bm k},X}({\bm r},t) \ ,
    \label{eq:mtdks}
\end{equation}
where the Kohn--Sham Hamiltonian is given by Eq.~(\ref{eq:hks}).  
The scalar and exchange correlation potentials also depend on $X$.  
The macroscopic current density ${\bm J}_{X}(t)$ at each $X$ is expressed using Eq.~(\ref{eq:current}) as
\begin{equation}
    {\bm J}_X(t) = {\bm J}[{\bm A}_X(t)](t) \ .
    \label{eq:mcurrent}
\end{equation}
By simultaneously solving Eqs.~(\ref{eq:multiscale}), (\ref{eq:mtdks}), and (\ref{eq:mcurrent}), we obtain the solution ${\bm A}_X(t)$ for a given initial pulse generated in the vacuum region in front of the thin film.  
At the beginning of the calculation, the Bloch orbitals $u_{n {\bm k},X}({\bm r},t)$, for all grid points of $X$ inside the thin film, were set to the ground-state solution.

For the incident laser pulse, the following time profile was adopted:
\begin{equation}
{\bm A}(t) = -\frac{cE_0}{\omega_0} \sin \left[ \omega_0 \left( t-\frac{T}{2} \right) \right] \sin^2 \left( \frac{\pi t}{T} \right) \hat z,
 (0 < t < T).
\end{equation}
where $E_0$ is the maximum amplitude of the electric field, $\omega_0$ is the fundamental frequency, and $T$ is the total pulse duration.
For the calculations, we set $T=20$ fs and $\hbar \omega_0 = 1.55$ eV.
The full width at half-maximum (FWHM) of the pulse is approximately 7 fs.
The initial vector potential at $t=0$ is expressed as
\begin{equation}
{\bm A}_X(t=0) = {\bm A} \left( - \frac{X}{c} \right).
\end{equation}

The reflectance $R$ and transmittance $T$ were evaluated as the ratio of the energies of the reflected and transmitted waves to the energy of the incident pulse.
The energies of the respective waves were evaluated at the final time of the propagation.
The absorbance $A$ was then given by $A=1-R-T$.

The multiscale Maxwell-TDDFT method has been successfully utilized for various nonlinear optical responses, including attosecond transient absorption spectroscopy \cite{Lucchini2016}, energy transfer from intense few-cycle pulsed light to dielectrics \cite{Lee2014, Sato2015, sommer2016}, propagation effects in high-harmonic generation \cite{Yamada2021, Yamada2023}, and the generation and propagation of coherent phonons \cite{Yamada2019-2}.

\subsection{Numerical detail}

\begin{table*}[t]
\centering
\begin{tabular}{ccccc}
\hline
    &    Al    & Graphite & Si   &  $\alpha$-quartz (SiO$_2$) \\
\hline \hline
Crystalline structure &  fcc  &hexagonal & diamond  &   rhombic  \\  
Lattice constants (\AA) & 4.049 & $a=6.696$, $b=4.254$, $c=2.456$ & 5.429 & $a=4.913$, $b=8.510$, $c=5.405$ \\
Number of atoms & 4 & 8 & 8 & 18 \\
Number of grid points & 30$^3$ & 32 $\times$ 20 $\times$ 16 & $32^3$ & 28 $\times$ 50 $\times$ 32 \\
k-points & 16 $\times$ 16 $\times$ 44 & adaptive \cite{Uemoto2021} & $12^3$ & $4^3$ \\
Functional & LDA \cite{Perdew1981} & LDA \cite{Perdew1981} & TB-mBJ \cite{Tran2009} & TB-mBJ \cite{Tran2009} \\
Film thickness (nm) & 50 & 200 & 200 & 200 \\
grid spacing for $X$ (nm) & 2.5 & 10 & 10 & 10 \\
Time step (as) & 0.5  &  0.5  &  0.5  & 0.5 - 0.3  \\
Number of time step & 60,000 & 60,000 & 60,000 & 60,000 \\
\hline
\end{tabular}
\caption{
Summary of the crystal structures, grid information, and exchange-correlation potential used for the four materials.
For all systems, a rectangular unit cell is used, with the $(a,b,c)$ axes set parallel to the $(x,y,z)$ directions. 
The light propagates along the $x$-direction and polarizes along the $z$-direction. The thin film is in the $yz$-plane. 
}
\label{tab-1}
\end{table*}

We utilized an open-source software package SALMON (Scalable Ab initio Light-Matter Simulator for Optics and Nanoscience) \cite{Noda2019, SALMON_web}, for which the authors are among the leading developers.  
In this code, the finite-difference method was used to solve for the electronic orbitals and electromagnetic fields.
The temporal evolution of the electron orbitals was determined using the Taylor expansion method \cite{Yabana1996}.

To solve the TDKS equation, we use orthogonal coordinate systems for four materials:, namely Al, Si, graphite, and SiO$_2$. 
The crystalline structures, crystalline parameters, and grid parameters of the unit cell and light propagation calculations, number of ${\bm k}$-points, time steps, number of time steps, and exchange-correlation potentials employed are summarized in Table~\ref{tab-1}.
We adopted an adiabatic approximation, utilizing the ground-state exchange-correlation potential in the time evolution calculation.
For Al and graphite, we employed a simple local density approximation (LDA) \cite{Perdew1981}.
For Si and SiO$_2$ that have finite band gaps, we employed the meta-GGA potential of Tran and Blaha \cite{Tran2009} (abbreviated TB-mBJ), which reproduces the bandgap energies of various dielectrics reasonably.

In first-principles optical response calculations, it is often necessary to use a large number of ${\bm k}$-points to obtain a convergent result.
This is particularly significant when long pulses with sharply defined frequencies are used.
Because multiscale Maxwell-TDDFT calculations require the extensive use of supercomputer resources and we employ a relatively short incident pulse of 7 fs at FWHM, we use moderately fine grid points in the ${\bm k}$-space.
We consider the results obtained here using the grid parameters shown in Table~\ref{tab-1} to be sufficiently accurate, at least for the discussion here.

\section{Results and discussion \label{sec:results}}

\subsection{Typical time-evolution calculation}

\begin{figure*}
    \includegraphics[width=0.88\linewidth]{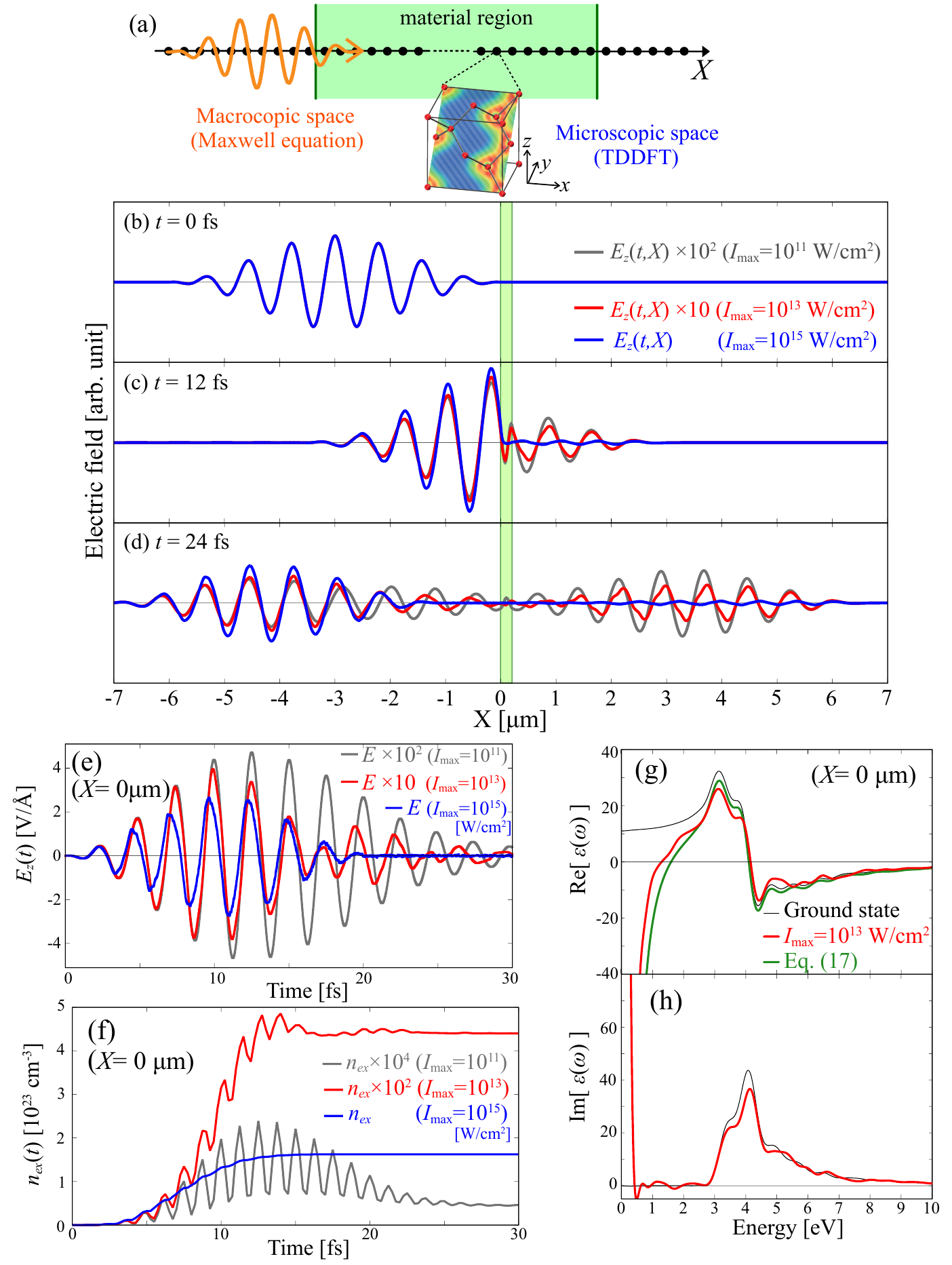}
    \caption{\label{fig:1} 
Overview of the multiscale Maxwell-TDDFT method for a pulsed light irradiation normally on a Si thin film of 200 nm thickness. 
(a) Grid system for the macroscopic coordinate $X$ and microscopic space that is prepared for each macroscopic grid point.
Snapshots of the electric field (b) at the initial time $t=0$; (c) at $t=12$ fs, when the pulse passes through the film; and (d) at $t=24$ fs when the pulse is separated into the transmitted and reflected waves.
Propagations of the three cases with different initial intensities are compared.
(e) Electric fields as functions of time and (f) number density of the excited electrons at the front surface ($X=0$) of the thin film for the three cases with different initial intensities.
(g) Real and (h) imaginary parts of the dielectric function at the front surface at $t=30$ fs.
See text for the detail.
    }
\end{figure*}

A typical calculation using the multiscale Maxwell-TDDFT method with a Si thin film as an example is shown in Fig.~\ref{fig:1}.
Fig.~\ref{fig:1}(a) shows the grid system used to calculate light propagation.
The Si thin film exists in the "material region" of the "macroscopic space.”
Because we set the film thickness to 200 nm and grid spacing for $X$ to 10 nm, as described in Table~\ref{tab-1}, the number of grid points for $X$ in the material region is 20.
At each grid point, indicated as "microscopic space,” electronic motion is calculated using the TDDFT.

Figs.~\ref{fig:1}(b--d) show the snapshots of the electric field $E_X(t) = -(1/c) d A_X(t)/dt$ at three different times: (b) at $t=0$, when the incident pulse is placed in the vacuum region in front of the film; (c) at $t=12$ fs, when the pulse passes through the film; and (d) at $t=24$ fs, when the pulse separates into transmitted and reflected pulses.
The three cases of pulse propagation with different maximum intensities of the incident pulse are compared in the panels shown in (b)–-(d).
Grey lines indicate the electric field for a weak incident pulse, where the maximum intensity is set to $I_{\rm max} = 10^{11}$ W/cm$^2$.
At this intensity, light propagation can be regarded as almost linear and be equivalently described by the dielectric function in the ground state.
Red lines indicate the electric field for a moderately intense incident pulse with $I_{\rm max} = 10^{13}$ W/cm$^2$. 
Blue lines indicate the electric field for a more intense incident pulse with $I_{\rm max} = 10^{15}$ W/cm$^2$. 
The central frequency of the pulse was set to 1.55 eV, which is below the direct bandgap of Si. The present calculation, which uses the TB-mBJ functional, yields 3.4 eV, which is close to the measured value.
The grey (red) lines are multiplied by a factor of 100 (10) such that the incident pulses of the three cases coincide with each other.
The difference between the gray and other colored lines shows the nonlinear effects on propagation.
As shown in the figure, nonlinear effects are more pronounced in the transmission than in the reflection.
For the case of $I_{\rm max}=10^{15}$ W/cm$^2$, the transmitted pulse almost disappeared owing to the strong nonlinear absorption, whereas the amplitude of the reflected pulse was the largest because of the plasma reflection that occurred at the front surface of the thin film.

In Fig.~\ref{fig:1}(e), the electric field at the front surface of the thin film ($X=0$) is plotted as a function of time.
The electric fields for the three cases with different incident intensities were compared.
The electric field at the front surface was given by the sum of the electric fields of the incident and reflected waves.
Although the reflected wave became stronger in the higher-intensity case, as shown in Fig.~\ref{fig:1}(d), the electric field at the front surface was weaker than that expected in the linear response, owing to the destructive interference between the incident and reflected waves.

Fig.~\ref{fig:1}(f) shows the number density of excited electrons in the unit cell at the front surface of the thin film, calculated using Eq.~(\ref{eq:nex}).
Again, the results for the three cases with different initial intensities were compared.
These were multiplied by the ratio of the intensity, $10^2$ ($10^4$), for $I_{\rm max}=10^{13}$ ($10^{11})$ W/cm$^2$.
If electronic excitations occur owing to linear absorption, the three lines should be identical.
Because the average frequency of the pulse (1.55 eV) is lower than the direct bandgap (3.4 eV), the excitation is dominated by multiphoton absorption, and the three lines show different behavior after multiplication.
At a weak intensity of $I_{\rm max} = 10^{11}$ W/cm$^2$, the oscillations are proportional to $E(t)^2$, which corresponds to the virtual electronic excitation.
When $I_{\rm max}=10^{15}$ W/cm$^2$, the number density becomes as high as approximately $1.6 \times 10^{23}$ /cm$^3$, which is close to the typical carrier density of metals.

Figs.~\ref{fig:1}(g) and (h) show the real and imaginary parts of the dielectric function, respectively, at $t=30$ fs.
At this time, most of the reflected and transmitted waves have exited the thin film, as shown in Fig.~\ref{fig:1}(d).
The dielectric function after pulse propagation was calculated using the procedure explained in Eqs.~(\ref{eq:probe})–(\ref{eq:eps_pump}), where the value of $t_0$ is set to $t_0=30$ fs.
Black lines indicate the dielectric function of Si in the ground state.
Red lines indicate the dielectric function with a pump pulse of $I_{\rm max}=10^{13}$ W/cm$^2$.
Green line in Fig.~\ref{fig:1}(g) shows the dielectric function of the following model adding the Drude contribution to that in the ground state.
\begin{equation}
\epsilon_D(\omega) = {\rm Re} \left[ \epsilon_{\rm gs}(\omega) \right] - \frac{4\pi n_{\rm ex}}{m^*\omega^2},
\end{equation}
where $n_{\rm ex}$ is the number density of the excited electrons at the front surface of the thin film at $t=30$ fs, and $m^*$ is the effective mass of the electron.
We use the value of $n_{\rm ex}$ obtained from Fig.~\ref{fig:1}(f) ($n_{\rm ex}=4.3 \times 10^{21}$ /cm$^3$) and set $m^*=0.18$ as employed in Ref.~[\onlinecite{Sokolowski2000}].
The green and red lines in Fig.~\ref{fig:1}(g) agree well indicating that the Drude model reasonably describes the change in the real part of the dielectric function, caused by the nonlinear electronic excitation.
A detailed discussion of this analysis is provided in Ref.~[\onlinecite{Sato2015}].

Because the medium is gradually excited by the electric field of the pulsed light, the dielectric function also changes over time from that in the ground state to that at $t=30$ fs.
Therefore, the reflectance of a pulse is not determined by the dielectric function shown in Figs.~\ref{fig:1}(g) and (h) but by the transient dielectric function.

\subsection{Systematic analysis: overview}

\begin{figure}
    \includegraphics[width=\columnwidth]{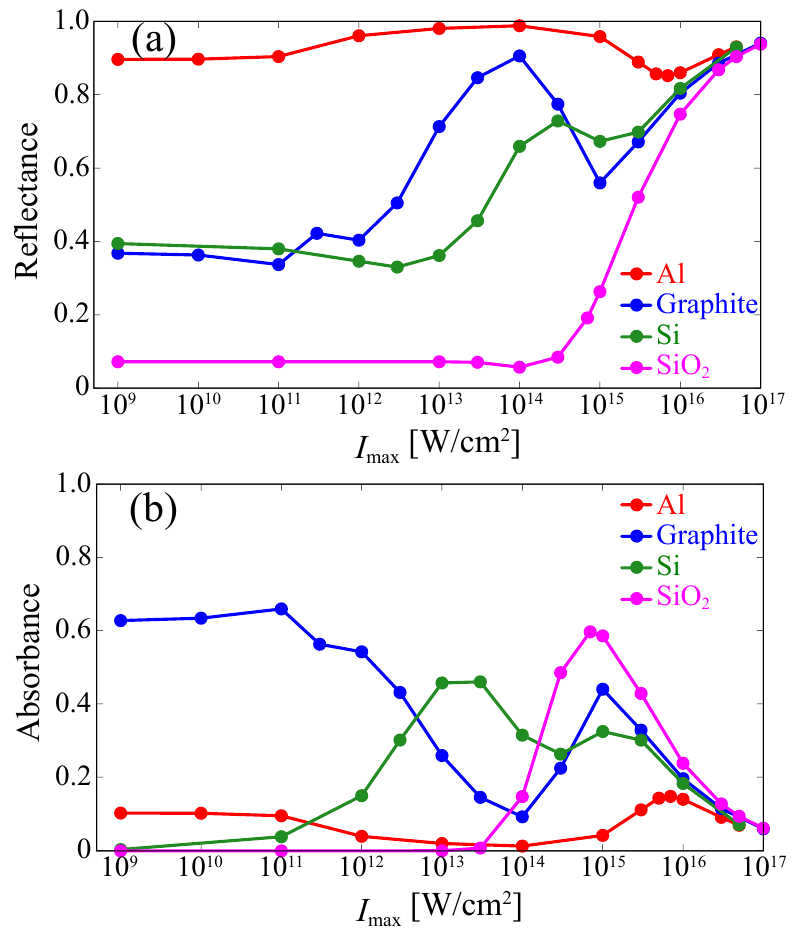}
    \caption{\label{fig:2} (a) Reflectance and (b) absorbance for a pulse irradiation of thin films of Al (red), graphite (blue), Si (green), and SiO$_2$ (purple) against the maximum intensity of the incident pulse.
    }
\end{figure}

We performed systematic calculations of light propagation through the thin films of four materials: Al (metal), graphite (semi-metal), Si (small-gap dielectric), and SiO$_2$ (wide-gap dielectric).
As in the previous subsection, linearly polarized light with a duration of $T=20$ fs (7 fs FWHM) and an average frequency of $\hbar\omega=1.55$ eV was used as the incident pulse for all calculations.
The film thickness was set to 200 nm for all materials except for Al, for which a thickness of 50 nm was used.
No transmitting wave was observed for the 50 nm thick Al thin film. 
Therefore, the results would not change when calculated for an Al thin film with a thickness of 200 nm.

We first present an overview of the reflectance and absorbance of the four materials when the maximum intensity of the incident pulse is changed from $I_{\rm max} = 10^9$ W/cm$^2$ to $10^{17}$ W/cm$^2$.
Figs.~\ref{fig:2} (a) and (b) show the results for the reflectance and absorbance, respectively.
First, we examined both sides of the weak and intense pulses.
For sufficiently weak incident pulses with $I_{\rm max} < 10^{10}$ W/cm$^2$, each material responds to light, as determined by its dielectric function in the ground state, metallic reflection in Al, and transparent propagation in SiO$_2$. 
For sufficiently intense pulsed light with $I_{\rm max} > 10^{16}$ W/cm$^2$, reflectance dominates in all materials, indicating that the materials lose their individuality and behave as a plasma of electrons.
A closer examination of the results presented in Fig.~\ref{fig:2} reveals that the change with the intensity of the light, the process by which a material loses its individuality and becomes a plasma, is different for each material.
In particular, the reflectance and absorbance of graphite and Si exhibit complex and non-monotonic changes in intensity. 

In the following subsections, we discuss the changes in the optical properties of each material in detail.
Before that, we discuss the adequacy and limitations of our theoretical and computational methods for light-matter interactions.
Our calculations are {\it ab initio} and do not include empirical parameters.
However, in practice, this is accompanied by significant approximations.
Since the TDKS equation (\ref{eq:tdks}) is a one-body Schr\"odinger-like equation, non-relativistic kinematics is assumed for the electronic motion.
The effects of the magnetic field on the electronic motion, which are significant at extremely high intensities, were ignored.
These approximations can be justified if the speed of electrons is much lower than that of light.
The light intensity in the atomic unit was $I=3.51 \times 10^{16}$ W/cm$^2$.
At approximately or above this intensity, we can expect the significance of the relativistic kinematics and Lorentz force due to the magnetic field for electronic motion.
However, the light field inside the material is different from that of the incident pulse.
For the sufficiently short pulse employed here, the electric field at the front surface of the thin film ($E_{\rm surface}$) is given by the sum of the electric fields of the incident ($E_{\rm incident}$) and reflected ($E_{\rm reflected}$) pulses ( $E_{\rm surface} = E_{\rm incident} + E_{\rm reflected}$).
Figure~\ref{fig:2} shows the plasma reflection dominates for extremely intense incident pulses.
As seen in Fig.~\ref{fig:1}(e), a strong cancellation occurs between the incident and reflected waves, such that the electric field inside the material ($E_{\rm surface}$) is much smaller than the electric field of the incident pulse ($E_{\rm incident}$).
In our calculations of the Si thin film, using an incident pulse with $I_{\rm max} = 10^{17}$ W/cm$^2$, the maximum electric field at the front surface was 12 V/m, corresponding to 0.23 in atomic unit.
Therefore, the present calculation that uses nonrelativistic kinematics and ignores the magnetic field can be justified even for the highest incident pulse of $I_{\rm max}=10^{17}$ W/cm$^2$ used in our calculations. 

The TDKS equation (\ref{eq:tdks}) is a one-body Schr\"odinger-like equation, where electron-electron interaction is treated only partially.
While part of the electron-electron interaction can be incorporated into the one-body electronic Hamiltonian of Eq.~(\ref{eq:hks}), such as the screening effect in the Hartree potential, electron–electron collision effects accompanying high-momentum transfer, such as avalanche and Auger recombination effects, are not included.
Because we fixed the atomic positions during the propagation calculation, no energy transfer from electrons to ions was considered.
These approximations can be justified in the present calculation using a very short incident pulse of 7 fs at FWHM.

\subsection{Systematic analysis: each material}

\subsubsection{Aluminum}

\begin{figure*}
    \includegraphics[width=0.95\linewidth]{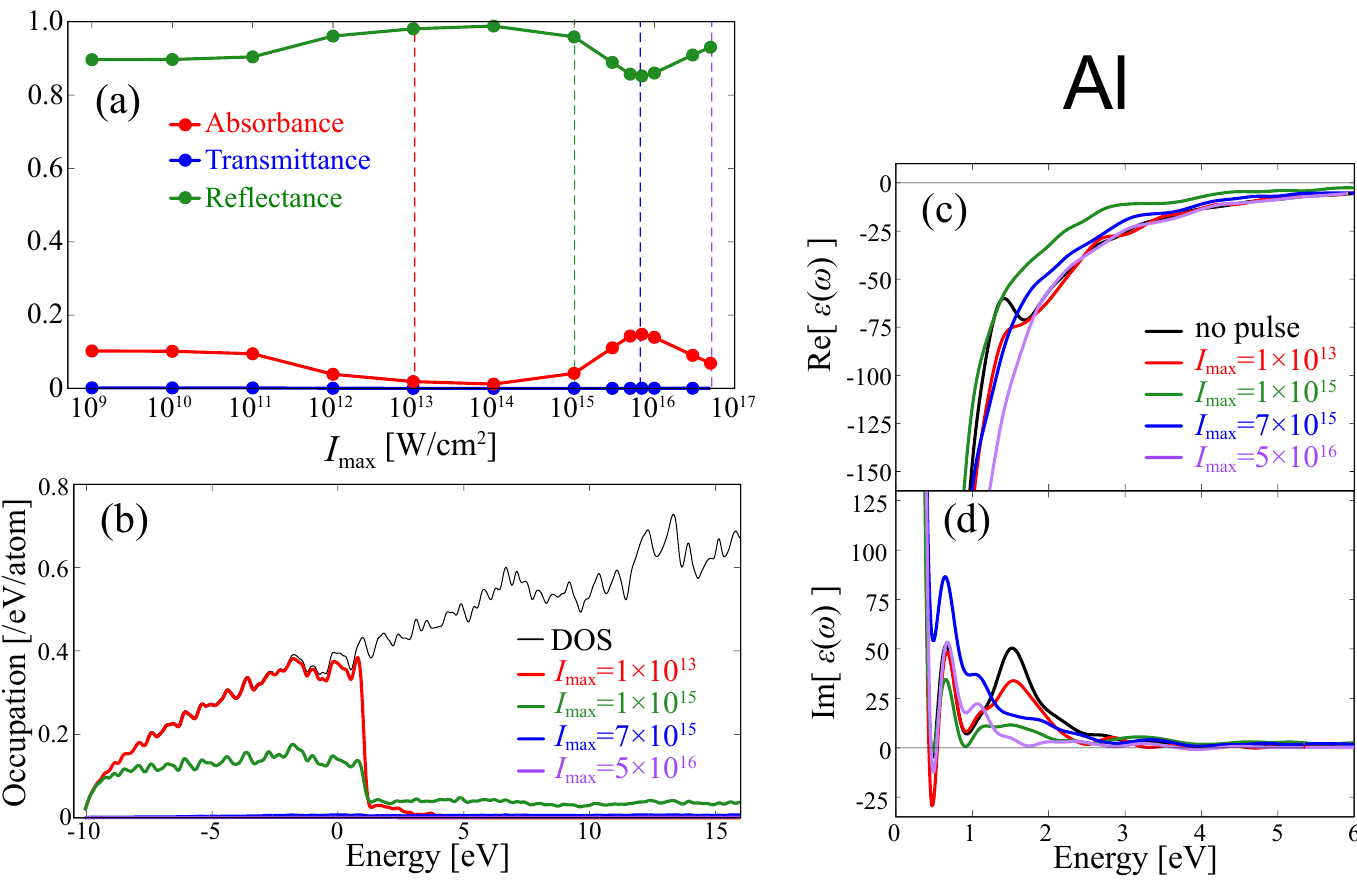}
    \caption{\label{fig:3} (a) Reflectance (green), absorbance (red), and transmittance (blue) of a pulsed light through an Al thin film with a thickness of 50 nm, shown against the maximum intensity of the incident pulse. Vertical dashed lines are drawn at the intensities where the lines are shown in (b--d) using the same colors. (b) Occupation distribution of electrons and (c) real and (d) imaginary parts of the dielectric function in the unit cell at the front surface of the thin film ($X=0$) at $t=30$ fs, for the cases of incident pulses with different intensities as indicated in the figure. The black line in (b) shows the density of states.
    }
\end{figure*}

First, we discuss the results of Al, which is a simple metal.
An adiabatic local density approximation was adopted for the calculation.
Fig.~\ref{fig:3}(a) shows the reflectance, transmittance, and absorbance of an Al thin film with a thickness of 50 nm against the maximum intensity of the incident pulse.
Fig.~\ref{fig:3}(b) shows the electron occupation distribution in the unit cell at the front surface of the thin film at $t=30$ fs for the four different incident intensities.
The total density of states is indicated by the black line.
Figs.~\ref{fig:3}(c) and (d) show the real and imaginary parts of the dielectric function, respectively.
In addition to the dielectric function in the ground state that is indicated as "no pulse,” the dielectric function in the unit cell at the front surface of the thin film at $t=30$ fs is shown for the four different incident pulse intensities.
The dashed vertical lines in Fig.~\ref{fig:3}(a) are drawn at the intensities at which the lines are shown in Figs.~\ref{fig:3}(b)--(d), using the same colors.

As seen in Fig.~\ref{fig:3}(a), reflection dominates at all intensities.
There was essentially no transmission in the film of 50 nm thickness at all intensity.
Slight absorptions appear at weak intensities ($I_{\rm max} < 10^{11}$ W/cm$^2$). 
As the intensity increased, the absorbance decreased, and the reflectance increased. Furthermore, the film exhibited almost complete reflection at an intensity of approximately $10^{13} \sim 10^{15}$ W/cm$^2$.
Subsequently, the absorbance increased to a peak value of 0.2 at around $10^{16}$ W/cm$^2$ and then decreased. 
Next, we consider the physical mechanisms of these variations in the reflectance and absorbance.

The slight absorbance at weak intensities ($<10^{11}$ W/cm$^2$) were caused by the interband transition, as indicated by the black line in Fig.~\ref{fig:3}(d), showing the finite imaginary part at $\hbar\omega=1.55$ eV.
We consider that an increase in the reflectance and a decrease in the absorbance that appear as the intensity increases (up to approximately $10^{13}$ W/cm$^2$) are caused by saturable absorption and are related to the change in the occupation distribution.
As seen in Fig.~\ref{fig:3}(b), a slight decrease in the occupation just below the Fermi level and slight increase just above the Fermi level can be observed at an incident intensity of $10^{13}$ W/cm$^2$. 
This small change in occupation is considered to be responsible for the reduction of the interband transition owing to the saturable absorption.  This hindrance of the absorption is caused by the decrease in valence electrons that can be excited by the linear absorption and by the increase in conduction electrons, which hinders the linear absorption by Pauli blocking.
The change in the occupation shown in Fig.~\ref{fig:3}(b) appears rather small at $10^{13}$ W/cm$^2$ after integration over the ${\bm k}$-space. Because linearly polarized light excites electrons in a specific direction, the change in occupation at a specific ${\bm k}$-value can be substantial, even if the change in the occupation distribution is small.

We consider that the increase in absorption at approximately $10^{15} - 10^{16}$ W/cm$^2$ is related to the significant change in the electron occupation distribution from a quantum Fermi distribution to a classical Boltzmann distribution.
As seen in Fig.~\ref{fig:3}(b), although the occupation shows a clear Fermi surface up to an incident intensity of $10^{13}$ W/cm$^2$, it becomes an intermediate state between the quantum and classical distributions at an intensity of $10^{15}$ W/cm$^2$. 
A classical Boltzmann distribution is observed at $7 \times 10^{15}$ W/cm$^2$.
The dielectric function at the front surface at $t=30$ fs is shown in Figs.~\ref{fig:3}(c) and (d). The real part of the dielectric function is dominated by a Drude-like negative contribution, irrespective of the incident intensity, and the imaginary part shows a complex change as the intensity increases.
The imaginary part of the dielectric function at $\hbar \omega = 1.55$ eV shows a decrease of $10^{13}$ W/cm$^2$ from that of the ground state.
It continued to decrease until $10^{15}$ W/cm$^2$ and then increased at $7 \times 10^{15}$ W/cm$^2$, which is consistent with the intensity dependence of the absorbance.
Therefore, the change in the absorbance is related to the change in the imaginary part of the dielectric function.
Thus, during the occupation change from the quantum Fermi distribution to the classical Boltzmann distribution, the imaginary part of the dielectric function increases, and the absorbance shows a maximum.

We now compare our results with the measurements in the literature.
Although the setups of reported experiments differ from the present calculation in many aspects, the comparison of calculations and experiments is considered important for clarifying the physical mechanism by which the light-matter interaction changes.
Price et al.~[\onlinecite{Price1995}] reported the absorbance of Al for an incident pulse with a wavelength and pulse duration of 800 nm and 120 fs, respectively, for a wide range of intensities, $10^{13} \sim 10^{18}$ W/cm$^2$.
It shows that the absorbance increases as the intensity increases, with a maximum value of approximately 0.3 at an intensity $3 \times 10^{14}$ W/cm$^2$.
Fig.~\ref{fig:3}(a) shows that the absorbance in our calculation increases above $10^{14}$ W/cm$^2$ and reaches a maximum value of approximately 0.2 at approximately $7 \times 10^{15}$ W/cm$^2$.
Therefore, the intensity of the maximum absorption shifted by a factor of 20, based on the measurement of [\onlinecite{Price1995}].
The pulse duration used in our calculation is about 7 fs at the FWHM, which is about 20 times shorter than that of the measurement.
If we compare the position of the maximum absorbance using the fluence of the incident pulse rather than the maximum intensity, our calculations and measurements will reasonably agree with each other.

In Ref.~[\onlinecite{Fisher2001}], measurements and theoretical analyses have been reported for the absorbance of a pulsed light from an Al surface with a wavelength and pulse duration 800 nm  and 20 fs, respectively, for the intensity range of $10^{11} \sim 10^{15}$ W/cm$^2$.
They observed a decrease of the absorbance from 0.15 to 0.1 when the intensity increased from $10^{11}$ W/cm$^2$ to $10^{13}$ W/cm$^2$ and attributed it to the decrease of the interband transition.
This trend of decrease in absorbance, along with its interpretation, is consistent with our results, although the minimum absorbance in the measurement (approximately 0.1) is larger than that in our calculation, which is much smaller than 0.1 and close to zero, as seen in Fig.~\ref{fig:3}(a).

Recently, the reflectance from Al surfaces was reported using pulses with wavelengths and durations of 800 nm and $15 \sim 100$ fs, respectively, for a fluence range of $0.05 \sim 10$ J/cm$^2$ \cite{Genieys2020}. 
The measurement shows a strong decrease in the reflectance as the fluence increases, with a reflectance as small as $0.2 \sim 0.3$ for a 15 fs pulse with approximately 5 J/cm$^2$, that is, approximately $3 \times 10^{14}$ W/cm$^2$. 
These results are inconsistent with our calculations or the previous measurements described above.
Currently, the cause of this discrepancy is not understood, and further studies are warranted.

\subsubsection{$\alpha$-quartz (SiO$_2$)}

\begin{figure*}
    \includegraphics[width=0.95\linewidth]{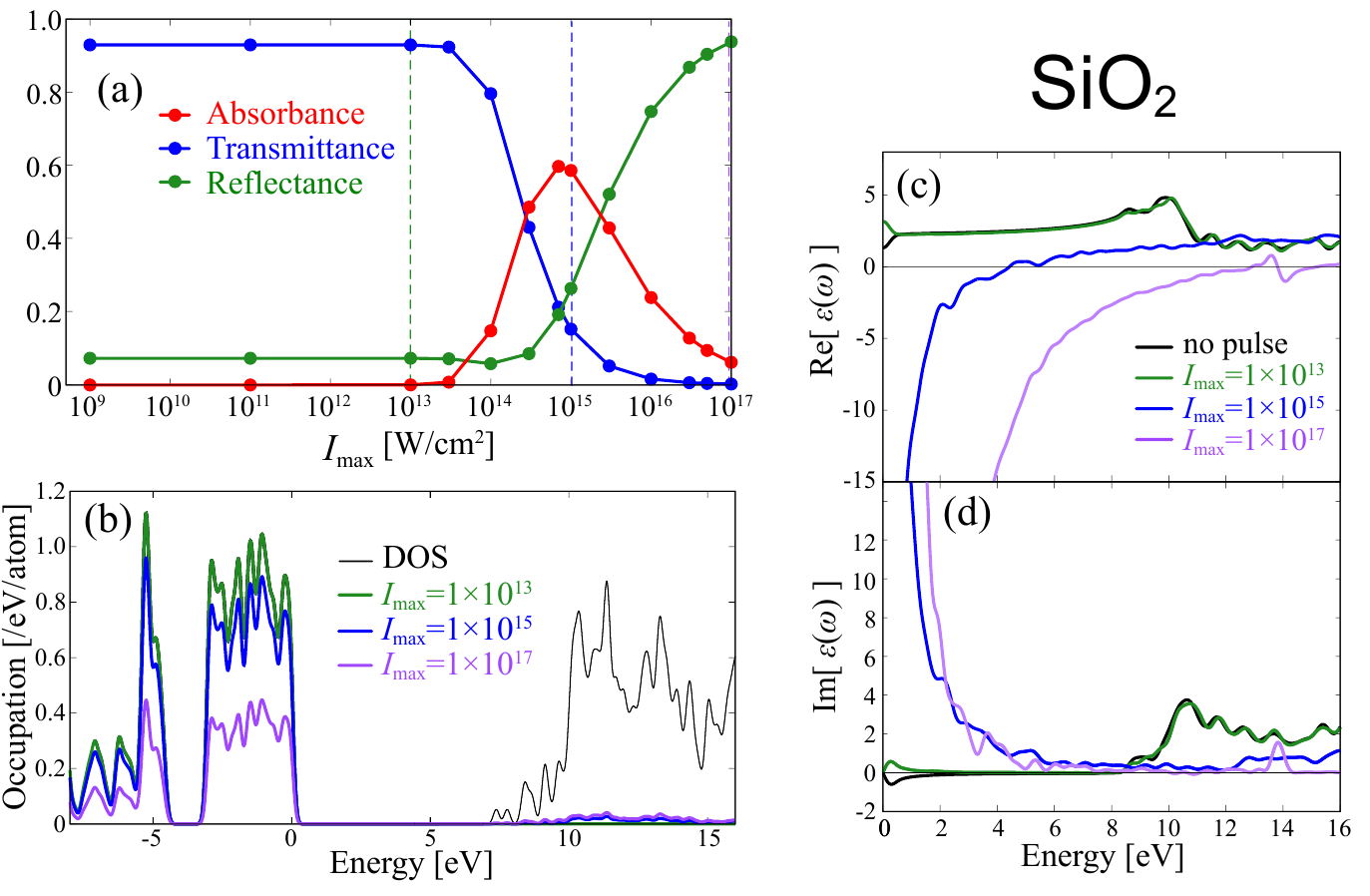}
    \caption{\label{fig:4} Same as Fig.~\ref{fig:3} except that this figure is for a thin film of $\alpha$-quartz with a thickness of 200 nm.
    }
\end{figure*}

Next, we discuss the results for $\alpha$-quartz (SiO$_2$), which is a typical wide-gap dielectric.
The calculation uses the meta-GGA potential of Tran and Blaha \cite{Tran2009}.
Using this potential, the calculated bandgap was 8.6 eV, which is close to the measured value.
For SiO$_2$, calculations for a smaller range of intensities have been reported using the multiscale Maxwell-TDDFT method \cite{Sato2015}.

Fig.~\ref{fig:4}(a) shows the reflectance, transmittance, and absorbance of SiO$_2$ thin film with a thickness of 200 nm against the maximum intensity of the incident pulse.
As shown in the figure, the high transmittance at a low intensity decreases rapidly when the laser intensity exceeds $10^{13}$ W/cm$^2$.
The decrease in transmittance is first compensated for by the absorbance, which shows a maximum of approximately 0.6 at approximately $10^{15}$ W/cm$^2$ and then decreases monotonically. 
The reflectance gradually increases above the mid-$10^{14}$ W/cm$^2$ and becomes dominant at and above $10^{16}$ W/cm$^2$.
The absorbance above the mid-$10^{13}$ W/cm$^2$ originates from a highly nonlinear multiphoton absorption process.
Because the bandgap of SiO$_2$ is 8.6 eV in the present calculation, absorption of more than five photons is required for the valence electrons to be excited into the conduction bands.

Fig.~\ref{fig:4}(b) shows the occupation distribution of an unit cell at the front surface of the thin film at $t=30$ fs.
At an incident intensity of $10^{15}$ W/cm$^2$, where the absorbance shows a maximum, the occupation distribution does not appear much different from that of the ground state, although a slight decrease is observed for the occupied orbitals.
Figs.~\ref{fig:4}(c) and (d) show the real and imaginary parts of the dielectric function, respectively. In addition to that in the ground state, shown by a black line and indicated as "no pulse,” those at the front surface of the thin film at $t=30$ fs are shown.
As the incident intensity increases, the real part of the dielectric function gradually receives a Drude-like negative contribution owing to electronic excitations.

In Fig.~\ref{fig:4}(a), a small dip is observed in the reflectance at approximately $10^{14}$ W/cm$^2$.
This behavior is related to the decrease in the real part of the dielectric function.
At an intensity of approximately $10^{14}$ W/cm$^2$, the real part of the dielectric function changes sign around the frequency of the laser pulse ($\hbar\omega=1.55$ eV).
This coincidence is known as the resonant condition, and it explains the dip in the reflection.
Above an intensity of $10^{14}$ W/cm$^2$, the real part of the dielectric function becomes negative at a frequency of $\hbar\omega = 1.55$ eV (Fig.~\ref{fig:4}(c)).
This implies that during pulse irradiation, the optical property of the material changes from dielectric to plasma, and the plasma-like reflection becomes dominant as the intensity of the incident pulse increases. 

To summarize, multiphoton absorption plays a dominant role in the intensity dependence in SiO$_2$. 
It first induces an increase in absorbance above an intensity of $10^{14}$ W/cm$^2$. 
It also creates free carriers in the conduction band, which add the Drude-like negative contribution to the real part of the effective dielectric function.
The change in the real part of the dielectric function induces an increase in the reflectance due to plasma reflection and a decrease in the absorbance above $10^{15}$ W/cm$^2$.
As the occupation distribution did not change significantly from the ground state at an intensity of $10^{15}$ W/cm$^2$, it did not play an important role in SiO$_2$, unlike the case of Al.

We compared the results of our calculations with the measurements and other theoretical analyses.
Price et al. [\onlinecite{Price1995}] reported the absorbance of SiO$_2$, as well as Al.
This shows that the absorbance was quite high, reaching a maximum value of approximately 0.9 at an intensity of approximately $2 \times 10^{13}$ W/cm$^2$ and then decreasing rapidly.
In our calculations, Fig.~\ref{fig:4}(a) shows that the absorbance reaches a maximum value of approximately 0.6 at approximately $7 \times 10^{14}$ W/cm$^2$ and then decreases.
Therefore, the intensity of the maximum absorption shifted by a factor of $30 \sim 40$ compared to the measurement.
Regarding the difference in the maximum value of the absorbance, there is still a transmittance of approximately 0.2 at the intensity that gives the maximum absorbance. This is due to our calculation being performed for a thin film of 200 nm.
This implies that the maximum absorbance can be as high as 0.8, which is the sum of the absorbance and transmittance, if calculations were performed for thicker films.
Regarding the shift in the intensity that gives the maximum absorbance, a similar shift was also observed for Al, as discussed in the previous subsection. This shift can be attributed to the differences in the pulse duration of approximately 7 and 120 fs FWHM in the present calculation and measurement, respectively.

Ziener et al. \cite{Ziener2002} reported the reflectance for incident pulses of wavelength and duration of 800 nm and 90 fs, respectively, from SiO$_2$ surface for incident intensities of $10^{13} \sim 5 \times 10^{17}$ W/cm$^2$.
In the measurement, the reflectance started to increase above $10^{14}$ W/cm$^2$ and rapidly increased to 0.8. 
Above $10^{17}$ W/cm$^2$, a rapid decrease in the reflectance was reported.
Considering the difference in the pulse lengths used in the experiments and calculations, we can conclude that our calculations agree reasonably well with the measurements.
Our calculations do not indicate any decrease in the reflectance at high intensities around $10^{17}$ W/cm$^2$.
The difference between the measurements and our calculation may be either because the pulse duration used in our calculation is much shorter than that used in the measurement, or the decrease in the measured reflectance originates from physical effects that were ignored in the present calculation.

Kaiser et.al. \cite{Kaiser2000} reported the theoretical analysis of absorption of intense pulses in SiO$_2$ using time-dependent Boltzmann equation that includes electron-electron collision effects.
They concluded that multiphoton absorption dominates in transparent solids, and the effects of the electron-electron collisions are small for a time scale of less than 100 fs.
This justifies our computational method, which ignores hard electron-electron collisions for pulses employed in the present work.

\subsubsection{Graphite}

\begin{figure*}[t]
    \includegraphics[width=0.95\linewidth]{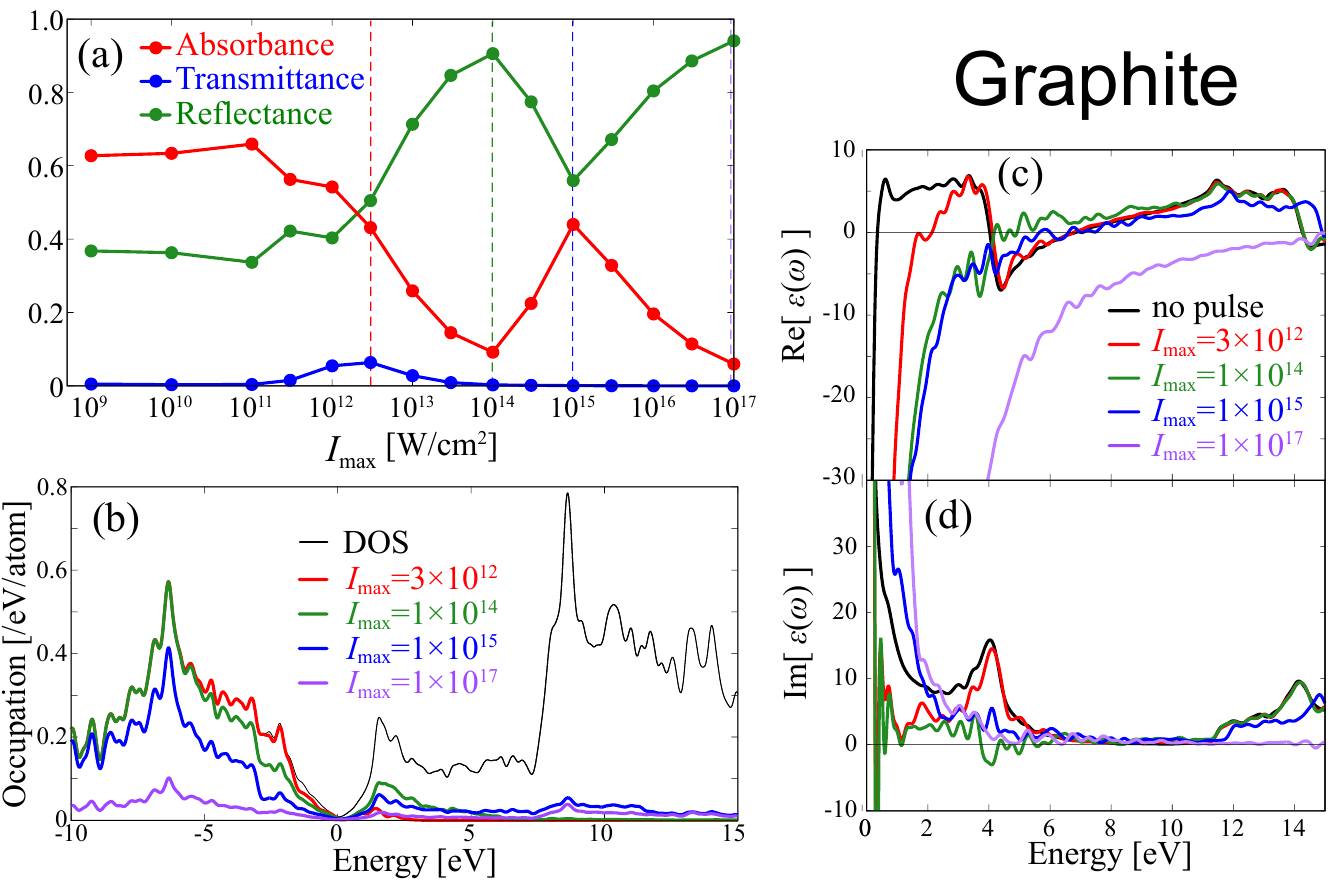}
    \caption{\label{fig:5} Same as Fig.~\ref{fig:3} except that this figure is for a graphite thin film of 200 nm thickness. }
\end{figure*}

In Fig.~\ref{fig:2}, the reflectance and absorbance of graphite and Si exhibit complex changes with incident intensity, contrary to those of Al and SiO$_2$. 
In this subsection, we discuss the results for graphite, which is a semi-metal.
The direction of light propagation was chosen to be perpendicular to the plane of the honeycomb lattice, and the electric field was in this plane.
It is noted that, in Table I, the axis perpendicular to the honeycomb lattice was chosen as the $a$-axis.
An adiabatic local-density approximation was used in the calculations.
Fig.~\ref{fig:5}(a) shows the reflectance, transmittance, and absorbance of a graphite thin film of 200 nm thickness against the maximum intensity of the incident pulse.
Fig.~\ref{fig:5}(b) shows the electron occupation distribution at the front surface of the thin film at $t=30$ fs for the four cases with different incident intensities.
Figs.~\ref{fig:5}(c) and (d) show the real and imaginary parts of the dielectric function, respectively. In addition to the dielectric function in the ground state indicated as "no pulse,” the dielectric function at the front surface at $t=30$ fs is shown for the four different incident intensities.

As seen in Fig.~\ref{fig:5}(a), the reflectance increases and absorbance decreases as the intensity increases. 
A small amount of transmittance appeared in the limited intensity region of approximately $10^{12} \sim 10^{13}$ W/cm$^2$. 
This was due to saturable absorption, which was also observed in Al.
It occurs due to a decrease in the valence electrons that can be excited by linear absorption and an increase in the conduction electrons, which hinders linear absorption by Pauli blocking.
Figs.~\ref{fig:5}(c) and (d) show that the real part of the dielectric function at $\hbar \omega = 1.55$ eV becomes negative and the absolute value increases to $10^{14}$ W/cm$^2$ as intensity increases, while the imaginary part decreases.
These changes are consistent with the observed decrease in absorbance and increase in reflectance in this intensity region.
In Ref.~[\onlinecite{Uemoto2021}], which is consistent with the present results, saturable absorption in graphite has been reported for a limited range of incident intensities using the multiscale Maxwell-TDDFT method.

The reflectance begins to increase at approximately $10^{11}$ W/cm$^2$ and reaches a maximum at an incident intensity of $10^{14}$ W/cm$^2$.
This was caused by the large negative value of the real part of the dielectric function (Fig.~\ref{fig:5}(c)).
Above $10^{14}$ W/cm$^2$, the reflectance decreased and reached a minimum at $10^{15}$ W/cm$^2$.
The absorbance shows a maximum at the same intensity.
Fig.~\ref{fig:5}(b) shows a change from almost full occupation below Fermi level to a classical Boltzmann distribution taking place across the intensity of $10^{15}$ W/cm$^2$.
The imaginary part of the dielectric function, shown in Fig.~\ref{fig:5}(d), exhibits a large value of $\hbar\omega = 1.55$ eV for this intensity.
From these observations, the appearance of the minimum reflectance and maximum absorbance at an approximate incident intensity of $10^{15}$ W/cm$^2$ is related to the change in the occupation distribution from a quantum Fermi distribution to a classical Boltzmann distribution.
A similar explanation has been provided for the absorption peak of Al at an incident intensity of $10^{16}$ W/cm$^2$.

Summarizing the results for graphite, three physical mechanisms play a role in intensity dependence.
Saturable absorption, which first appears at an intensity of $10^{12} \sim 10^{13}$ W/cm$^2$, reduces the absorbance and increases the transmittance.
Above this intensity region, electronic excitations induce a negative contribution to the real part of the effective dielectric function, and plasma reflection becomes dominant.
Upon further increasing the intensity, a drastic change in the occupation distribution from a quantum Fermi distribution to a classical Boltzmann distribution occurred at an intensity of $10^{15}$ W/cm$^2$, and the absorbance (reflectance) reached the maximum (minimum).
Upon further increasing the intensity, the plasma reflection became dominant.

We compared our calculated results with the measurements.
Breusing et al. \cite{Breusing2009} reported pump-probe measurements of graphite flakes of $20 \sim 30$ nm thickness using a 7 fs pulse with a central frequency of 1.55 eV and 
showed an increase in the transmittance of 20 \% at a pump intensity of $5 \times 10^{10}$ W/cm$^2$.
Winzer et al. \cite{Winzer2012} reported the theoretical and experimental results of saturable absorption for a graphite thin film with 60 layers (approximately 20 nm).
For pulses with frequency and duration of 1.5 eV and 56 fs, respectively, absorption saturation was observed at $5.7 \times 10^{10}$ W/cm$^2$.
Under such conditions, the measured values did not coincide with the present calculations.
The intensities of the incident pulses that showed an increase in the transmittance were lower than those in our calculations, and the amount of transmittance was small in our calculations.
We consider these differences to originate from the difference in the film thickness between the measurements ($20 \sim 30$ nm) and our calculation (200 nm).
In Ref.~[\onlinecite{Uemoto2021}], we report the calculated results for a graphite thin film of 50 nm thickness using the multiscale Maxwell-TDDFT method.
An increase in transmittance was found to occur around the incident intensity of $10^{11}$ W/cm$^2$, and the amount of increase in transmittance was approximately 0.4 at approximately $10^{12}$ W/cm$^2$, which were in better agreement with the measurements.

\subsubsection{Silicon}

\begin{figure*}[t]
    \includegraphics[width=0.95\linewidth]{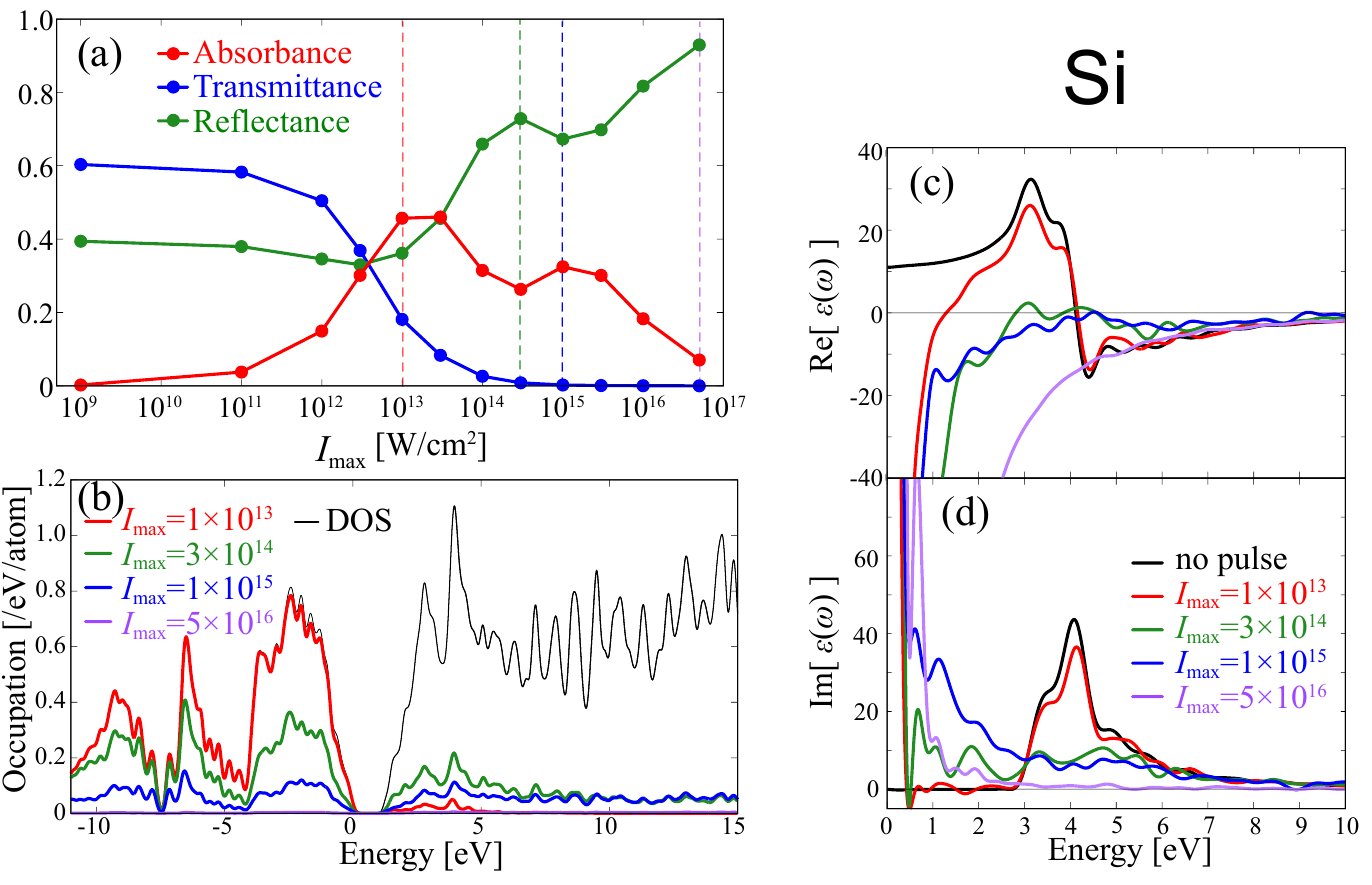}
    \caption{\label{fig:6} Same as Fig.~\ref{fig:3} except that this figure is for a Si thin film of 200 nm thickness.  }
\end{figure*}

Finally, we present the results obtained for Si, which is a small-gap dielectric.
The calculation used the meta-GGA potential of Tran and Blaha \cite{Tran2009}.
The bandgap calculated using this potential was 3.4 eV, which is close to the measured value.
The first application for the irradiation of the bulk surface of Si was carried out in Ref.~[\onlinecite{Yabana2012}], where the multiscale Maxwell-TDDFT method was developed.
Recently, high harmonic generation from Si thin films was reported using the same method \cite{Yamada2021, Yamada2023}.

Fig.~\ref{fig:6}(a) shows the reflectance, transmittance, and absorbance of Si thin film with a thickness of 200 nm against the maximum intensity of the incident pulse.
Silicon should have weak absorption at $\hbar \omega = 1.55$ eV in the linear response owing to the one-photon absorption accompanying phonons. 
However, in the present calculation, such processes are ignored because we fix the atomic positions at the equilibrium positions in the primitive cell.
Thus, the thin film was transparent at low intensities.
In Fig.~\ref{fig:6}(b), the electron occupation distribution at the front surface of the thin film at $t=30$ fs is shown for the four incident intensities.
Figs.~\ref{fig:6}(c) and (d) show the real and imaginary parts of the dielectric function, respectively. 
In addition to the dielectric function in the ground state, denoted as "no pulse,” dielectric function at the front surface at $t=30$ fs is shown for the four different incident intensities.

As the light intensity increased, the transmittance gradually decreased, and the absorbance increased owing to multiphoton absorption. 
The reflectance decreased slightly at an intensity of $10^{12} \sim 10^{13}$ W/cm$^2$ and then increased.
This can be explained by the behavior of the real part of the dielectric function at $\hbar\omega = 1.55$ eV as shown in Fig.~\ref{fig:6}(c).
The real part of the dielectric function approaches zero at an intensity of $10^{13}$ W/cm$^2$ and becomes negative above this intensity.
This behavior of the real part of the dielectric function causes a slight decrease in the reflectance at an intensity of $10^{12} \sim 10^{13}$ W/cm$^2$ and then an increase.
A similar behavior was observed for SiO$_2$.

Above $10^{13}$ W/cm$^2$, reflectance increases, while absorbance and transmittance decrease.
However, changes in the incident intensity do not follow a monotonic pattern but exhibit complex behavior. 
The reflectance shows a maximum and then a minimum at approximately $10^{14} \sim 10^{15}$ W/cm$^2$, whereas the absorbance shows the opposite trend. 
Occupational distributions are shown in Fig.~\ref{fig:6}(b). The occupation of the valence orbitals exhibits a large change in this intensity region.
The imaginary part of the dielectric function, shown in Fig.~\ref{fig:6}(d), exhibits a significant enhancement at $\hbar\omega=1.55$ eV, at an incident intensity of $10^{15}$ W/cm$^2$.
We consider the complex change of the reflectance and absorbance as the increase of the intensity to be related to the rapid change of the occupation distribution, from mostly occupied valence orbitals to a classical Boltzmann-like occupation.
A similar mechanism for producing the maximum absorbance was also observed in aluminum and graphite, as shown in Figs.~\ref{fig:3}(a),(b), and(d) and \ref{fig:5}(a),(b), and(d). 
Thus, we consider that such enhanced absorption linked to the drastic change in occupation distribution is a universal phenomenon in the interaction between intense and ultrashort pulsed light.

In summary, multiphoton absorption first induces a nonlinear absorption and then decreases the transmittance above an intensity of $10^{11}$ W/cm$^2$.
Electronic excitation contributes negatively to the real part of the effective dielectric function.
As the intensity increases, plasma reflection gradually becomes significant and induces a decrease (increase) in absorbance (reflectance) above an intensity of $10^{13}$ W/cm$^2$.
Upon further increasing the intensity, a drastic change in the occupational distribution from a quantum distribution to a classical Boltzmann distribution occurred at an intensity of $10^{14} \sim 10^{15}$ W/cm$^2$.
The absorbance (reflectance) showed the maximum (minimum) value.
Plasma reflection dominated upon further increase of intensity.

We compared our calculated results with the measurements.
Hulin et al.~[\onlinecite{Hulin1984}] reported a change in the reflectance of the Si surface for a pulse of duration and wavelength of 100 fs and 620 nm, respectively.
As the intensity of the pulse increased, a small decrease of approximately 5\% occurred in the reflectance, followed by an increase of approximately 50\% when the intensity increased by a factor of 10.
These behaviors are consistent with our calculations, although the pulse durations and frequencies are different.
In their analysis, Drude model was used to describe the changes in optical properties caused by electronic excitations.
The Drude model includes two parameters: carrier density and collision time.
A number density of free carriers between $5 \times 10^{21}$ cm$^{-3}$ and $2 \times 10^{22}$ cm$^{-3}$ was adopted.
This value is consistent with our calculations.
As shown in Fig.~\ref{fig:1}(f), the number density of excited electrons at $t=30$ fs is $4.4 \times 10^{21}$ cm$^{-3}$ at an intensity of $I_{\rm max} = 10^{13}$ W/cm$^2$.
This intensity is close to the intensity at which reflectance is minimized.
The collision time is set as $3 \times 10^{-16}$ s. 
The authors state that this very short collision time originates from electron-hole interactions.
A small electron relaxation time has also been adopted in the literature \cite{Sokolowski2000, Medvedev2010}.
The mechanism of the collision, specifically the electron-hole interaction effect, is taken into account in our multiscale Maxwell-TDDFT method. When solving the TDKS equation in real time, the scattering of electrons by ions, which induces interactions between electrons and holes, is fully taken into account.

Pump-probe measurements of the reflectance of Si and theoretical analyses have been reported in Ref.~[\onlinecite{Sokolowski2000}].
In their theoretical analysis, a minimum of reflectance was predicted at the excited electron density of about $7 \times 10^{21}$ cm$^{-3}$, and a rapid increase of reflectance was predicted above this value of the excited electron density.
These estimates are consistent with our calculations and those of Ref.~[\onlinecite{Hulin1984}].
A similar behavior of reflectance against intensity was reported in Ref.~[\onlinecite{Danilov2015}] using pulses with wavelengths and duration of 800 nm and 100 fs, respectively.
However, Ref.~[\onlinecite{Riley1998}] reported that no clear evidence of enhanced reflectance was observed at high intensities when using an incident pulse with a wavelength and duration 750 nm and 60 fs, respectively.

\section{Summary}

In this study, we present a systematic theoretical analysis of the propagation of intense ultrashort pulsed light through thin films of various materials.
We employed the first-principles multiscale Maxwell-TDDFT method, in which microscopic electronic motion and macroscopic light propagation are coupled using a coarse-graining approximation.
The reflectance, transmittance, and absorbance of a thin film with a thickness of $50 \sim 200$ nm were calculated for four materials, namely Al (metal), graphite (semi-metal), Si (small-gap dielectric), and SiO$_2$ (wide-gap dielectric), with different band structures and optical properties.
Linearly polarized pulsed light with an average frequency of 1.55 eV, a pulse duration of 7 fs at FWHM, and maximum intensities of $10^9 \sim 10^{17}$ W/cm$^2$ normally irradiated the thin films.

From the systematic calculations of the four materials, we found the following characteristics in nonlinear optical response:
While the optical responses to weak light are described by the dielectric function, which varies significantly depending on the material, the response to very intense light is dominated by plasma reflection, regardless of the material.
The process of plasma reflection becomes dominant as the intensity increases non-monotonically. 
As the light intensity increases, perturbative nonlinearities first appear. 
Multiphoton absorption appears in dielectrics, where the bandgap energy is larger than the frequency of light, whereas saturable absorption appears in metals and semi-metals where the linear excitation occurs at the frequency of the light.
In dielectrics, a small minimum reflectance appears when the real part of the effective dielectric function that incorporates the effects of excited carriers crosses zero.
The maxima in absorbance and minima in reflectance appear for all four materials at an intensity of $10^{15} \sim 10^{16}$ W/cm$^2$.
In Al, graphite, and Si, this originates from a significant change in the occupation distribution from a quantum-fully occupied distribution of valence orbitals to a classical Boltzmann distribution.
The imaginary part of the effective dielectric function is enhanced during the change in the distribution.
In SiO$_2$, although maximum absorption appears at a similar intensity, it is not related to the drastic change in the occupation distribution but to the interplay between multiphoton absorption and plasma reflection.
Above an incident intensity of $10^{16}$ W/cm$^2$, all the materials showed plasma reflection, characterized by large negative values in the real part of the effective dielectric function.

Note that the calculations are {\it ab initio} and employ no empirical parameters.
For weak light, our formalism of the multiscale Maxwell-TDDFT method results in ordinary macroscopic electromagnetism with the dielectric functions given by TDDFT. 
Because the TDKS equation is solved in real time and no perturbative approximation is employed for the electronic motion, the nonlinearities in the electronic response are fully considered in the calculation.
Because we employed a three-dimensional grid system in solving the TDKS equation, there was no truncation in the spatial degrees of freedom, that is, the unoccupied orbitals were not truncated.

However, numerous physical effects are ignored in our framework.
Because we assume fixed atomic positions at the equilibrium positions in the primitive cell, the thermal effects related to ionic motion and energy transfer between electrons and ions are not included.
For the very short pulse with a duration of a few tens of femtoseconds used here, the energy transfer from electrons to ions may not be important.
Another serious approximation in the present TDDFT calculations for electronic motion is the adiabatic approximation, in which the exchange-correlation functional in the ground state is employed for time-dependent motion.
As we are essentially solving a one-body TDKS equation, only limited many-body correlation effects are incorporated.
Because we calculated Hartree and exchange-correlation potentials at each time step, dynamic screening effects were included through the time-dependent Hartree potential.
However, the electron-electron collision effects accompanying large momentum or energy transfer were not included.
In particular, the avalanche mechanism, which is important for pulses with durations of a few tens of femtoseconds or longer, was not included.
Notably, despite these many approximations, the present calculation plausibly captures the variation in the optical response in a wide intensity region for various materials, reaching the plasma reflection, all within the framework of first-principles calculations.

\begin{acknowledgements}
This research was supported by JST-CREST under the grant number JP-MJCR16N5, MEXT Quantum Leap Flagship Program (MEXT Q-LEAP) (Grant number: JPMXS0118068681), and JSPS KAKENHI (Grant number: 20H2649). 
Calculations were performed on the Fugaku supercomputer with the support of the HPCI System Research Project (Project ID: hp230130) and Wisteria at the University of Tokyo, with the support of the Multidisciplinary Cooperative Research Program in CCS, University of Tsukuba.
\end{acknowledgements}

\appendix

%


\end{document}